\documentclass{article}
\usepackage[utf8]{inputenc}
\usepackage[yyyymmdd]{datetime}
\usepackage{amssymb}
\usepackage{url}
\usepackage{tabularx}
\usepackage{tikz}
\usepackage{wasysym}
\usepackage{flowchart}
\usepackage{breakurl}
\usepackage{pgf-umlsd}
\usetikzlibrary{shapes,arrows.meta,chains}

\title{Certifying Digitally Issued Diplomas}
\author{Geoffrey Goodell\\University College London\\\texttt{g.goodell@ucl.ac.uk}}
\date{\small \textit{This Version: \today}}

\definecolor{ourcol}{rgb}{0.7, 0.2, 0.5}
\definecolor{mfill}{rgb}{0.99, 0.9, 0.6}
\definecolor{mborder}{rgb}{0.8, 0.6, 0.4}
\definecolor{col1}{rgb}{0.7, 0.2, 0.5}

\definecolor{col2}{rgb}{0.5, 0.2, 0.9}

\definecolor{col3}{rgb}{0.5, 0.5, 0.5}

\definecolor{col4}{rgb}{0.0, 0.5, 0.7}

\newcommand{\cz}[1]{\textit{\textbf{#1}}}

\newcolumntype{L}[1]{>{\raggedright\arraybackslash}p{#1}}
\newcolumntype{C}[1]{>{\centering\arraybackslash}p{#1}}
\newcolumntype{R}[1]{>{\raggedleft\arraybackslash}p{#1}}

\begin{document}

\maketitle

\begin{abstract}

We describe a protocol for creating, updating, and revoking digital diplomas
that we anticipate would make use of the protocol for transferring digital
assets elaborated by Goodell, Toliver, and Nakib~\cite{goodell2022}.  Digital
diplomas would maintain their own state, and make use a distributed ledger as a
mechanism for verifying their integrity.  The use of a distributed ledger
enables verification of the state of an asset without the need to contact the
issuing institution, and we describe how the integrity of a diploma issued in
this way can persist even in the absence of the issuing institution.

\end{abstract}

\section{Digital certificates}

A digital certificate comprises two parts: a datagram and a cryptographic signature over the datagram from an issuer.  A digitally issued Diploma contains this certificate along with a unique (and possibly empty) set of updates that accumulate over time.  In the certificate, the datagram \textit{may} contain:

\begin{itemize}

\item the time of issuance;

\item the identity of the student (the “diploma holder”);

\item the identity of the issuing institution or department (the “diploma issuer”), possibly accompanied by a certificate (chain) authorising that institution or department to issue diplomas;

\item the identity of the specific person who awarded the diploma, possibly accompanied by a certificate (chain) authorising that person to award diplomas;

\item the specific nature and qualifications of the diploma (this can also be achieved by having different signing keys for different kinds of diplomas);

\item the identity of the party or set of parties that can update or revoke the diploma; and

\item the expiration date, if any.

\end{itemize}

The aforementioned fields, if present, \textit{should} (after decoding, as necessary) contain human-readable information, although in some cases that might not be necessary.  The issuer \textit{must} choose a specific validator node participating in the digitally issued Diploma programme to be the ``integrity provider'' for this digitally issued Diploma.  In all cases, the datagram \textit{must} contain:

\begin{itemize}

\item a public key (or the output of a deterministic one-way hash function thereof) specifying the unique, one-time private key that will be used to create this digitally issued Diploma,

\item a public key (or the output of a deterministic one-way hash function thereof) specifying the unique, one-time private key that can update (or revoke) this digitally issued Diploma in the future, and

\item a unique, deterministic reference (for example, the hash) to a recent comprehension of transactions, which we call a ``tethering point'', maintained by the integrity provider (note: this \textit{may} be known in advance or \textit{may} be requested from the integrity provider immediately before creating the digitally issued Diploma).

\end{itemize}

From time to time, the digitally issued Diploma \textit{may} be updated by the holder of the unique, one-time private key specified in the most recent update, if one exists, or the original certificate, otherwise.  Each update \textit{should} (after decoding, as necessary) contain human-readable information indicating the nature of the update and \textit{must} contain the public key (or a deterministic one-way hash output thereof) denoting the unique key that can update (or revoke) the digitally issued Diploma in the future.  The update \textit{may} contain the specification of a new integrity provider.

To issue a digitally issued Diploma or to perform an update, the issuer \textit{must} send the following data to the chosen integrity provider for this digitally issued Diploma:

\begin{itemize}

\item the output of a deterministic one-way hash function applied to the initial certificate or update,

\item the public key corresponding to the unique, one-time private key that will be used to create this digitally issued Diploma or to create this particular update, and

\item a signature of the initial certificate or update using the corresponding private key.
An update can be used for routine key rotation or to delegate control of a digitally issued Diploma to another party.  A revocation is one kind of update, and it can be reversed in a future update.

\end{itemize}

The diploma issuer is responsible for maintaining all of its signing keys, including all of the one-time keys that are denoted for updates.  If any of these keys are lost or compromised, then the integrity of the affected digitally issued Diplomas \textit{should} be considered compromised as well.  The process for alerting relying parties about a process failure on the part of a diploma issuer that has led to one or more compromised digitally issued Diplomas is outside the scope of the validation process; in general it is assumed that process failures are made public knowledge via a timely notification mechanism.  One way to implement this mechanism is via periodic (or just in time) validation of the certificates issued to diploma issuers themselves, possibly using the same process used for issuing and updating digitally issued Diplomas.

\section{Certificate validation}

The validity of a digitally issued Diploma depends upon its authenticity, integrity, and uniqueness as of a (sufficiently recent) moment in time:

\begin{itemize}

\item The authenticity of a digitally issued Diploma is determined by evaluating the signature of its initial certificate and an affirmation of trust in the diploma issuer and the process for managing process failures.

\item The integrity of a digitally issued Diploma is determined by verifying that the initial certificate, all of its updates, and no additional updates are present in a version of history that is authorised (signed) by its integrity provider (and its designated successors, if any) from the moment of issuance through the specified moment in time.

\item The uniqueness of a digitally issued Diploma is determined by verifying that the version of history is consistent with the commitments made by the integrity provider (and its designated successors, if any) to the ledger.

\end{itemize}

It is assumed that the diploma holder has been provided a copy of the certificate, along with any updates that have accrued from time to time, by the issuer (and its designated successors, if any).  If this is not the case, then it will not be possible to validate the certificate.

It is possible for the diploma holder, or, equivalently, anyone to whom the diploma holder shares the diploma and its updates (the ``prover''), to verify the authenticity, integrity, and uniqueness of the diploma (as of a moment in time) given the following additional information:

\begin{itemize}

\item a sufficiently recent ledger entry, signed by the relevant set of ledger validators, and

\item a ``proof of provenance''.

\end{itemize}

Since the ledger is assumed to be public (that is, there is a suitable API for retrieving the ledger contents from validator nodes), the only challenge for the prover is to obtain a proof of provenance.  The proof of provenance contains the following information:

\begin{itemize}

\item Evidence that the hash of the initial certificate has been comprehended by a commitment made by the (initial) integrity provider to the ledger, and

\item Evidence that zero or more updates have been made to this digitally issued Diploma, each by the public key specified in the previous update, if one exists, or the initial certificate, otherwise, and

\item Evidence that, up to a sufficiently recent point in time, there have been no updates beyond the last update, if one exists, or the original certificate, otherwise, and

\item Evidence that all updates, if any, have been comprehended by commitments made by the (corresponding) integrity provider(s) to the ledger.

\end{itemize}

A prover \textit{may} request a proof of provenance from any validator node participating in the digitally issued Diploma programme.  This is done by sending:

\begin{itemize}

\item the tethering point for this digitally issued Diploma,

\item the public key used to create this digitally issued Diploma, and

\item the public keys used to create the successive updates of this digitally issued Diploma, if any.

\end{itemize}

\section{Roles}

\cz{Role of the diploma holder:} The diploma holder \textit{should} maintain a safe copy of the digitally issued Diploma, including any updates that it receives.  The diploma holder \textit{may} request proofs of provenance from time to time.  To assert that the diploma holder has a valid diploma, the prover \textit{must} provide the digitally issued Diploma to the relying party.  The prover \textit{may} provide a proof of provenance to the relying party as well, although the relying party \textit{may} also request a proof of provenance from a validator.

Role of the diploma issuer: For every certificate that it issues, the diploma issuer \textit{must} choose a validator node participating in the digitally issued Diploma programme to serve as the integrity provider.  The diploma issuer \textit{may} choose different integrity providers for different certificates.  For digitally issued Diplomas issued by a diploma issuer to remain valid, the diploma issuer \textit{must} maintain the security of its valid one-time keys for the most recent update of each diploma that it has issued (and for which the expiration date, if any, has not yet been reached).

The diploma issuer \textit{should} rotate these keys from time to time via the update mechanism, and such rotations \textit{should} make use of appropriate improvements in public-key cryptography.  The diploma issuer \textit{should} also maintain the security of the institutional keys that it uses to sign the one-time keys that it uses to sign the initial certificates.  The diploma issuer \textit{should} appoint an appropriate successor in the event that it can no longer continue to perform its required tasks; otherwise, its failure to do so \textit{must} be recognised by the mechanism used to handle process failure.  The diploma issuer \textit{must} notify the mechanism used to handle process failure in the event of any errors in process or in the event that a key has been compromised. 

\cz{Role of the integrity provider:} An integrity provider \textit{must} maintain an up-to-date history, perhaps as a series of signed hashes, each comprehending the previous hash and any new transactions that have taken place since the creation of the previous hash.  An integrity provider \textit{must} commit its most recent version of history to ledger from time to time and \textit{should} make such commitments in a timely manner.  An integrity provider \textit{must} periodically provide copies of the set of transactions, comprising public keys as well as the opaque hashes of certificates or updates, for each moment in its history, to all validator nodes participating in the digitally issued Diploma programme.

The integrity provider does not receive copies of digitally issued Diplomas or their contents.  The integrity provider \textit{may} charge a fee for registering a new certificate or an update.  The integrity provider \textit{may} (or \textit{may} not) distinguish between initial certificates and updates (the hashes are opaque, so without an explicit mechanism, the integrity provider might not be able to know the difference).

An integrity provider \textit{should} rotate its keys from time to time for the purpose of externalising its commitments to the ledger, and such rotations \textit{should} make use of appropriate improvements in public-key cryptography.  An integrity provider \textit{should} appoint an appropriate successor in the event that it can no longer continue to perform its required tasks; otherwise, its failure to do so \textit{must} be recognised by the mechanism used to handle process failure.  An integrity provider \textit{must} notify the mechanism used to handle process failure in the event of any errors in process or in the event that a key has been compromised.

\cz{Role of validator nodes participating in the digitally issued Diploma programme:} All validator nodes participating in the digitally issued Diploma programme of an association \textit{should}:

\begin{itemize}

\item On request, serve as integrity provider for a digitally issued Diploma.

\item On request, and given appropriate information, provide a proof of provenance for a digitally issued Diploma.

\item Notify the association if it is unable to provide a proof of provenance for a particular digitally issued Diploma because of having received insufficient information from the corresponding integrity provider.

\end{itemize}

Validator nodes \textit{may} charge a fee for validation service.

Validator \textit{should} pay the association for their role in the issuance and updating of digitally issued Diplomas, and from the generation of proofs of provenance.  This can be done either by remitting some fixed or proportional share of the proceeds of these services, or by remitting some proportional share of the size of transaction metadata that they share with other participants in the digitally issued Diploma programme, or both.

\section{Privacy considerations}

The management of diplomas is oblivious: They do not appear on the ledger, and the validator nodes never receive diplomas or their contents directly.  However, if the pre-image of a digitally issued Diploma is leaked, it might be possible for a validator node to confirm its existence.

If a prover does not provide a proof of provenance to the relying party, then the relying party seeking to validate a digitally issued Diploma \textit{must} request a proof of provenance.  The process of requesting proofs of provenance generates metadata, including not only the identity of the requesting party but also the one-time public keys involved in the certificate and its updates.  Although those keys might not reveal the specific identity of the diploma holder, they can be recognised via successive requests from relying parties.  In some circumstances, the diploma holder or the relying party might enjoy better privacy if the diploma holder requests proofs of provenance directly from a validator node, and shares them with all of the relying parties, even though the diploma holder might reveal its own metadata to the validator node when it does.  To address this risk, an issuer might consider furnishing the diploma holder with multiple diplomas or furnishing the diploma holder with new diplomas including all applicable updates on request.

\section{Security considerations}

Diploma issuers and integrity providers have a responsibility to manage their private keys, which includes appropriate procedures for secure storage and rotation.  These responsibilities are coupled with a mechanism for handling process failure.  This mechanism \textit{may} involve verifying certificates for diploma issuers and integrity providers via the same mechanism used to manage and validate digitally issued Diplomas.

Integrity providers are required to share their transaction metadata with each other.  In principle, they \textit{may} fail to do so, perhaps because of an interest in reducing their fees to the association, or possibly because of a misconfiguration or error.  In such cases, it is assumed that other validator nodes will be able to request the transactions.  However, if the integrity provider has failed without providing the complete set of transaction metadata, then it might not be possible to validate the digitally issued Diplomas for which it served as integrity provider, particularly if no sufficiently recent proofs of provenance have been generated for those digitally issued Diplomas.

It is possible for a validator node to isolate digitally issued Diplomas from other transactions that it intends to keep private.  This can be done by aggregating the digitally issued Diplomas to a single, public hash entry within a larger, generally private, set.  It is also possible to aggregate private transactions into a single hash entry within the set of hashes for digitally issued Diplomas.  In this manner, validator nodes are required to share metadata about digitally issued Diplomas but are not required to share similar metadata about other transactions.

Since the ledger and transaction metadata stores contain no data about diplomas themselves, but only hashes, then as long as the assumption of the irreversibility of cryptographic hashes holds, and preimage data about diplomas are not compromised, then the ledger and transaction data stores can safely be stored in perpetuity.

Transaction metadata \textit{must} be held for the lifetime of digitally issued Diplomas, which can, in principle, be eternal.  Eternal data storage is a well-known problem in archival science and, in a peer-to-peer network, can be addressed via distributed document preservation systems such as LOCKSS, which does not require the use of a distributed ledger.  A document preservation system \textit{may} be combined with a process for periodically copying archive contents to non-electronic offline storage for additional robustness.

\section*{Acknowledgements}

The author expresses thanks to Dr Jörn Erbguth and Prof Tomaso Aste, who have
provided motivation, insight, and valuable feedback that underpin the ideas
expressed in this article.

\end{document}